\vskip 0.1cm


\vfill\eject

{\bf I. Introduction and motivation for the experiment}

For highly excited interacting many-body systems the independent
particle picture has very little validity when the mean spacing
between the many-body energy levels is much smaller than the spacing of the 
single-particle levels [1,2]. For high excitations, the 
interaction leads to a quick decay of the single-particle as well
as of collective modes [3] which are not eigenstates
of the total Hamiltonian of the system. This decay results in a formation
of highly complicated many-body configurations.
Each of these many-body states is characterized by uniform
occupation of all accessible parts of phase space and sharing of energy
between many particles of the system. The characteristic time
for the formation of such ergodic, independent of the initial conditions,
many-body states is given by the inverse spreading width,
$\tau_{erg}=\hbar/\Gamma_{spr}$ [3]. 

Consider a highly excited many-body system, whose spectrum obeys
Wigner-Dyson statistics, for the time interval
$t\gg\tau_{erg}$. Does ergodicity of all {\sl individual} many-body
eigenstates necessarily imply that their superposition 
is incoherent
random superposition? Can superposition of spatially extended ergodic 
modes of a highly
excited many-body system
produce localized or non-equilibrium non-ergodic patterns?
 In the absence of a theory 
 for phase randomization in an isolated (disconnected from a heat bath)
systems one may conventionally rely on the hypothesis emerging from
the foundations and modern developments of the random matrix theory (RMT)
of highly excited many-body systems. This hypothesis implies that the 
energy relaxation, i.e. the mere formation of ergodic individual 
many-body configurations, is a sufficient condition for a phase 
randomization between these ergodic eigenstates [3]. If true, this
conjecture should validate universal applicability of the RMT for the
energy interval $\Delta E\leq \Gamma_{spr}$ and for the time interval
$t\geq\tau_{erg}$, accordingly.

Consider the decay of a highly excited many-body system with strongly
overlapping resonances, $\Gamma\gg D$, where $\hbar/\Gamma\gg\tau_{erg}$
is the average life-time and $D$ is the mean level spacing of the system.
This regime, $D\ll\Gamma\ll\Gamma_{spr}$, is known as a regime of 
Ericson fluctuations [4,5] for the decay of equilibrated nuclear, 
atomic and
molecular systems and in coherent electron transport through 
nanostructures [3]. Suppose that RMT 
universally applies for $t\gg\tau_{erg}$. This implies absence of 
correlations between transition amplitudes (partial width amplitudes), 
known as Bethe's
random signs hypothesis, for the decay of different
ergodic states to either the same or different quantum micro-channels
[3]. Consider, e.g., a strongly 
dissipative heavy-ion collision (DHIC) characterized by a high
intrinsic excitation energy ($\geq 15$ MeV) of the double (deformed) 
intermediate system. Since for nuclear systems
$\Gamma_{spr}\simeq 5$ MeV and, for DHIC, $\Gamma\simeq 100$ keV [6]
we deal with the decay of a superposition of ergodic strongly 
overlapping
($\Gamma\gg D$) many-body configurations. Then the RMT
hypothesis, $\tau_{deph}\leq\tau_{erg}$ with $\tau_{deph}$ being
the phase randomization (dephasing) time between ergodic states,
implies that the cross sections for the DHIC, summed over a very large
number of partial cross sections, corresponding to different
micro-states of the reaction fragments, should show a smooth energy 
dependence
with the characteristic energy variation $\geq\Gamma_{spr}\simeq 5$ MeV. 
In contrast, experimental studies [7-9] present overwhelming evidence
for the persistence of rapid ($\simeq 100$ keV) energy oscillations
in the cross sections for DHIC. This manifests itself in the correlations 
between different transition amplitudes indicating that the phase 
randomization between individual ergodic configurations should be a much
slower process than energy relaxation ($\tau_{deph}\gg\tau_{erg}$)
in sharp contrast with the RMT hypothesis. 

In attempting to interpret the
 non-self-averaging of excitation function oscillations
in DHIC one faces a non-straightforward task of realization of 
Wigner's dream [10], namely to modify RMT by
taking into account level-level and channel-channel correlations 
between the transition
amplitudes. Such a possible modification has been presented
in Refs. [11,12] in terms of spontaneous coherence 
and slow phase randomization in highly excited many-body systems. 
While RMT
develops ``a new statistical mechanics'' (Dyson) of, by purpose, 
fully equilibrated
finite systems, the work [14,15] discusses critical and 
non-equilibrium phenomena in finite highly excited many-body systems.

It has been found [11-13] that a precondition for micro-channel 
correlations (MC) in complex quantum collisions is $\tau_{deph}\gg\tau_{erg}$.
Physically, $\tau_{deph}$ sets up a new time scale for quantum many-body 
systems. For times $t <\tau_{deph}$, the RMT ceased to apply even though
$t\simeq\hbar/\Gamma\gg \tau_{erg}$. Since the physical picture 
[11-13]
for the MC is in a sharp contrast with the RMT and the theory of
quantum chaotic scattering [3] it is highly desirable to have an
additional independent test of the approach [11-13]. Such a possibility
does indeed exist. It has been argued [14] that the 
spontaneous origin of MC should result in the cross sections for DHIC
being sensitive to an infinitesimally small perturbation. 

There is convincing evidence that the effects of complexity and
stochasticity in nuclear systems are shared by other microscopic
and mesoscopic many-body systems [3]. Therefore the spontaneous MC
and the extreme sensitivity should be expected for other complex 
quantum collisions, e.g., atomic, molecular, and atomic cluster
collisions.

The discussion of Ref. [14] has not taken into
account different distributions of electro-magnetic fields,
defects etc. within different independently prepared target foils. 
How might the consideration [14] apply in
the presence of such differently distributed ``target-environmental'' 
perturbations within different targets?

Consider a simple case of spinless reaction partners in the entrance 
and exit channels. A generalization for the case of the reaction
partners having intrinsic spins is straightforward. 
Then the measured cross section, per a single target nucleus and for a 
fixed single exit micro-channel $\bar b$
(microscopic states of the reaction products), is given by
$$
\sigma_{\bar b}(E,\theta )=(1/{\cal N})\sum_{j=1}^{\cal N}
\sigma_{\bar b}^{(j)}(E,\theta),
\eqno{(1)}
$$
where
$$
\sigma_{\bar b}^{(j)}(E,\theta)=|f_{\bar b}^{(j)}(E,\theta)|^2.
\eqno{(2)}
$$
Here index $(j)$ labels individual target nuclei participating
in the collision, whose number is 
${\cal N}\gg 1$, 
and $f_{\bar b}^{(j)}(E,\theta)$ is the amplitude
of a collision involving $(j)$ target nucleus. 
The difference between
$f_{\bar b}^{(j)}(E,\theta)$ with different $(j)$ originates from
a nonuniform distribution of ``target-environmental'' perturbations. 
This introduces different local perturbations, $V_j\neq V_i$, in the 
 purely nuclear Hamiltonian $H$
of highly excited nuclear molecules created in the collision of the
incident ion with different $(j\neq i)$ target nuclei.
We evaluate the strength of the ``target-environmental'' perturbations
to be of the order of the atomic electron effects [14] in DHIC.
We employ the perturbation theory [14] and use the 
decomposition $f_{\bar b}^{(j)}(E,\theta)=f_{\bar b}(E,\theta)+
\delta f_{\bar b}^{(j)}(E,\theta)$, where $f_{\bar b}(E,\theta)$
is the collision amplitude in the absence of the 
``target-environmental'' perturbations.
We also drop the incoherent sum
$(1/{\cal N})\sum_{j=1}^{\cal N}
|\delta f_{\bar b}^{(j)}(E,\theta)|^2$. 
This sum is about fourteen orders of magnitude smaller than 
$\sigma_{\bar b}(E,\theta)$.
We obtain
$
\sigma_{\bar b}(E,\theta)=|F_{\bar b}(E,\theta)|^2-
 |(1/{\cal N})\sum_{j=1}^{\cal N} \delta f_{\bar b}^{(j)}(E,\theta)|^2
\to |F_{\bar b}(E,\theta)|^2,
$
where $|(1/{\cal N})\sum_{j=1}^{\cal N} \delta f_{\bar b}^{(j)}(E,\theta)|^2
\leq 10^{-14}|F_{\bar b}(E,\theta)|^2$, and
$F_{\bar b}(E,\theta)$ is the collision amplitude corresponding
to the Hamiltonian $(H+v)$ with $v=(1/{\cal N})\sum_{j=1}^{\cal N}V_j$.

It is reasonable to assume that a distribution of the local 
``target-environmental'' perturbations $V_j$ is random throughout the target.
This means that $\delta f_{\bar b}^{(j)}(E,\theta)$ with different $(j)$
have random phases. In this case we have 
$$
|F_{\bar b}(E,\theta)-
f_{\bar b}(E,\theta)|\sim (1/{\cal N})^{1/2}
|\delta f_{\bar b}^{(j)}(E,\theta)|\sim (1/{\cal N})^{1/2}10^{-7}
|f_{\bar b}(E,\theta)|,
$$
 where 
we used the estimate $|\delta f_{\bar b}^{(j)}(E,\theta)|\sim
10^{-7}|f_{\bar b}(E,\theta)|$ from Ref. [14].

Suppose we perform two independent measurements with two different
targets. The ``target-environmental'' perturbations,
$V_j$ in the first target and ${\tilde V_j}$ in the second one, 
are different. The cross sections are given by the
different amplitudes, $F_{\bar b}(E,\theta)$ and 
${\tilde F}_{\bar b}(E,\theta)$, corresponding to different Hamiltonians,
$(H+v)$ and $(H+{\tilde v})$, accordingly.
 Then we have 
$|F_{\bar b}(E,\theta)-{\tilde F}_{\bar b}(E,\theta)|\sim 
(1/{\cal N})^{1/2}
10^{-7}|F_{\bar b}(E,\theta)|$.
 Therefore one does not expect a
detectable difference for the cross sections measured with two different
targets. Indeed, such a detectable difference does not occur
 if one considers  
$\sigma_{\bar b}(E,\theta)$ for a single fixed $\bar b$ independently
from the cross sections for the decay to other $\bar b^\prime\neq \bar b$
micro-channels.
However, as suggested in Ref. [14], the situation may change drastically
for the cross sections summed over very large number of exit micro-channels. 
This is the case for DHIC where
 the collision products have high excitation energies and
the measured cross section, $\sigma(E,\theta)=
\sum_{\bar b}\sigma_{\bar b}(E,\theta)$, is the sum over very large number 
of micro-channels, $N_{\bar b}\gg 1$.

The above consideration suggests that
 the spontaneous MC might lead to up to 100$\%$
difference between the non-self-averaging oscillating components
of the cross sections for DHIC for two measurements with different targets.
The key element in the interpretation of the spontaneous coherence,
non-self-averaging and extreme sensitivity in complex quantum collisions
is introduction of the infinitesimally small off-diagonal 
MC between
different {\sl model} transition amplitudes which couple {\sl model} 
single-particle states
(Slater determinants) of the quasi-bound IS and the continuum
 states [11-14]. It has been argued that
the limit of the vanishing of this infinitesimally small correlation
properly supplemented by the limit of the infinite dimensionality of the
Hilbert space does not destroy correlation between different 
{\sl physical} transition amplitudes which couple the many-body 
configurations
of the IS and the continuum states.
As a result, 
the highly-excited thermalized (${~}\hbar/\Gamma\gg\tau_{erg}$) 
matter displays
 coexistence of two distinct phases. The decay of the disordered phase
is associated with the $\Delta S_{\bar b}^J$-matrix [11], 
where $J$ is the total spin
of the IS, and, thereby, with the amplitude $\Delta F_{\bar b}(E,\theta)$
which is a linear combination of $\Delta S_{\bar b}^J$ with different $J$.
Since $\Delta F_{\bar b}(E,\theta)$ with different  
$\bar b\neq\bar b^\prime$ do not correlate,
this disordered phase does not contribute 
to the MC producing the stable reproducible self-averaging, i.e. energy 
smooth, background in cross sections.
The non-self-averaging, i.e. micro-channel correlations, and 
sensitivity originate from decay of the ordered phase 
corresponding
to the micro-channel independent $\delta S^J$-matrix [13,14] and, thereby,
the micro-channel independent $\delta F(E,\theta)$.
It is this micro-channel independent $\delta F(E,\theta)$ which
is so sensitive and, therefore, non-reproducible due to the 
 spontaneous origin of the MC so that 
$|\delta F(E,\theta)-\delta {\tilde F}(E,\theta)|\sim |\delta F(E,\theta)|
\sim |F_{\bar b}(E,\theta)|$, where
$\delta F(E,\theta)$ and $\delta {\tilde F}(E,\theta)$ correspond to 
different targets with different distributions of 
``target-environmental'' perturbations.

It follows from the above consideration and Ref. [14] that the 
non-self-averaging energy oscillating component of the cross section
is determined by the ``target-environmental''
perturbations averaged over the whole target and not by the micro-channel
differences unless these differences produce  
different perturbations of the Hamiltonian of hot
double intermediate system. Indeed the detailed energy dependence
of the non-self-averaging energy oscillating component of the cross section
is determined by the amplitude $\delta F(E,\theta)$ which does not depend
on the micro-channel indices. The case when the micro-channel differences
in the {\sl entrance} channel do produce different perturbations
of the Hamiltonian of the intermediate system is considered in Ref. [14].

Pictorially, the sensitivity of the $\delta S^J$-matrix and 
$\delta F(E,\theta)$
resembles the sensitivity
of the direction of the spontaneous magnetization vector, below the Curie
point, to the 
direction of an infinitesimally small external magnetic field.
 
 {\bf II. Experimental method}
 
In order to test the sensitivity
two independent measurements of excitation functions
for the strongly dissipative collision for the same reaction system of 
$^{19}$F+$^{93}$Nb have been carried out at the
China Institute of Atomic Energy (CIAE), Beijing. In these measurements,
the $^{19}$F$^{8+}$ beam was provided by the HI-13 tandem accelerator.
The beam incident energies were varied from 102 to 108 MeV in steps 
of 250 keV. For both measurements the same accelerator 
parameters and the same electronic and acquisition systems were 
selected. 
The same two sets of gas-solid ($\Delta E-E$) telescopes,
with a charge resolution $Z/\Delta Z\geq 30$ and an energy resolution
$\leq 0.4$ MeV, were set 
at 38$^{\circ }$ and 53$^{\circ }$. 
The $\Delta E$ detector is an ionization chamber filled with P10 gas
at a pressure of 103 mb, the residual energy $E$ is deposited in a 
Si position sensitive detector with a thickness of 1000 $\mu$m, a size
of 8$\times$47 mm and a marked position resolution of 0.5 mm. The solid
angles of the two telescopes are 1.80 msr and 2.62 msr, respectively.
Count rates in the experiment were less than 10 counts per sec. so that
a pile-up problem did not occur.

In Fig. 1 we present a typical ($\Delta E-E$) scatter-plot obtained at
$E_{lab}(^{19}$F)=100.25 MeV. It is seen that the projectile-like 
fragments from the $^{19}$F+$^{93}$Nb reaction can be separated.
For the F fragments direct and quasi-elastic processes constitute
the major contribution into the cross section. For the Ne fragments
there was no a sufficient statistics. Therefore we restrict our
analysis to the cross sections of the N and O products   
of the $^{19}$F+$^{93}$Nb DHIC. 

To avoid a possible effect of the carbon build up in the target,
we analyse events with $E_{lab}($N)$\geq$50 MeV and 
$E_{lab}($O)$\geq$55 MeV for $\theta_{lab}=38^\circ$, and
with $E_{lab}($N)$\geq$40 MeV and 
$E_{lab}($O)$\geq$40 MeV for $\theta_{lab}=53^\circ$ (see Fig. 1).
In Fig. 2 we present a ($\Delta E-E$) scatter-plot for the fragments 
from the $^{19}$F+$^{12}$C reaction at $E_{lab}(^{19}F)=100.25$ MeV. 
Our measurement shows that the cross
sections are negligible for the N and O outgoing energies $\geq 45$ MeV 
for $\theta_{lab}=38^{\circ }$, and  $\geq 40$ MeV for 
$\theta_{lab}=53^{\circ }$. We also note that, for 
$E_{lab}(^{19}F)=108$ MeV and
$\theta_{lab}=53^{\circ }$, the production of the N and O fragments
with the outgoing energy $\geq 39$ MeV
in the $^{19}$F+$^{12}$C reaction is kinematically forbidden.
Since the energies of the N and O yields
in our measurements  $\geq$50 MeV for $\theta_{lab}=38^{\circ }$ and
$\geq$40 MeV for $\theta_{lab}=53^{\circ }$ 
 we conclude that the carbon build up does not produce uncontrolled
errors and does not affect our data for the 
cross sections of the N and O products   
of the $^{19}$F+$^{93}$Nb DHIC. 

In Fig. 3 we present, as an example, a typical energy spectrum of 
the dissipative N yield for $E_{lab}(^{19}$F)=103.25 MeV and 
$\theta_{lab}=38^{\circ }$ produced in the $^{19}$F+$^{93}$Nb DHIC
 for one run
in the second experiment (see Fig. 5). Therefore the counting rate
for the correspondent data in Fig. 6 is about as twice as higher
than that in Fig. 3.

In Fig. 4 we present angular distributions for the N outgoing 
fragments measured at $E_{lab}(^{19}$F)=100 MeV and 105 MeV.
The angular distributions are strongly forward peaked due to
the major contribution of direct fast reaction processes at the
forward, near grazing, angles.

In the two measurements we used different, independently 
prepared, self-supporting $^{93}$Nb target foils with the 
 thickness $\simeq$70 $\mu g/cm^2$.
Both the target foils were produced by the sputtering method. The 
 thickness of each of the two foils was determined by
the spectrophotometry. It was found that the difference in thickness
of the two foils $\leq$5 $\mu g/cm^2$.
This difference results in different stopping energy losses  
in the two different targets. However, this itself should not affect
reproducibility of the cross sections since 
 this difference in stopping energy losses 
$\sim 15$ keV is smaller than the energy
spread $\sim 50$ keV in the beam and additional energy spread 
$\sim 150$ keV in the target.

 Absolute cross sections
were not determined, though great care was taken to ensure no spurious
sources of oscillations were introduced into the relative cross sections.
The stability of the beam direction was controlled as follows:
(i) TV monitor screen 
was used before each energy step to check and correct the position  
of the beam spot on the target. (ii) Two silicon 
detectors were placed at $\theta_{lab}=\pm 12^{\circ}$. (iii) The 
 beam charge was collected using a Faraday cup placed
at $\theta=0^{\circ}$ and was compared with the counting rates of 
the silicon detectors. The data were normalized both with respect to the
count rates of each of the silicon detectors and the integrated beam 
current. All the three normalizations produced the relative
cross sections, for each individual experiment, which agree 
 within the statistical errors, $1/N^{1/2}$, where $N$ is a count rate.
We have taken 5 repeat points
(one repetition for 5 different energies measured)
for the first experiment (target) and 21 repeat points
(one repetition for 21 different energies)
for the second experiment (target). 
Before to repeat each point
the TV monitor screen was used to check and correct a position of
the beam spot. All the repeated points
demonstrated the reproducibility, within the statistical errors,
for both individual experiments (targets). This reproducibility 
is demonstrated in Fig. 5 for the two runs in the second experiment. 
Such a reproducibility for the two runs for the same targets 
indicates 
that no damages of the targets, which could bring about uncontrolled 
spurious effects, occurred in our experiments.
All the above procedures indicate that the systematic uncertainties
do not seem to be present and the data errors can be evaluated as 
statistical only.

{\bf III. Experimental results}

The cross sections $\sigma (E)$ for the products N and O in the 
$^{19}$F+$^{93}$Nb
DHIC are presented in Fig.6, where the error bars are statistical only.
Although Fig. 6 presents energy integrated yields over the wide,
$\sim 25$ MeV, ranges of the
dissipative spectra (i.e. these yields
are summed over huge number of different final micro-channels of
the highly excited collision products)
the characteristic non-self-averaging oscillating structures of the
excitation functions in DHIC can be visually identified.

Taking into account that an energy resolution
 of our  ($\Delta E-E$) telescopes $\leq 0.4$ MeV, from Fig. 3 we
find that possible cross section energy variations due to the lower
energy cut-off are $<1\%$ for the dissipative N yield at
$\theta_{lab}=38^{\circ }$. From the energy spectra of the
dissipative O yield we found that the lower energy 
cut-off also produces negligible, $<1\%$, cross section energy 
variations
for the O reaction products at
$\theta_{lab}=38^{\circ }$.

From Fig. 6 we notice that, for some incident energies, 
the cross sections measured for two different
target foils are different. A statistical significance
of this non-reproducibility is discussed in Section IV.

For $E_{lab}(^{19}$F)=105 MeV,
the total excitation energy of the double intermediate system is $E$=87 MeV.
It consists of the deformation energy $E_{def}$, the rotational energy 
$E_{rot}$ and the
intrinsic excitation energy $E^\ast$. The deformation energy 
is mainly given by the Coulomb energy of the two touched ions
which yields $E_{def}\simeq 43$ MeV. We calculate the average rotational
energy for a rigid body moment of inertia of the two touched ions
with the average angular momentum $\bar J=(J_{cr}+J_{gr})/2$, where
$J_{cr}$ and $J_{gr}$ are the critical and the grazing angular momenta,
respectively. In our case, $J_{cr}=40$ and $J_{gr}=53$ in $\hbar$ units.
We have $E_{rot}$=27 MeV and $\hbar\omega=1.2$ MeV, where $\omega$ is
the average angular velocity of the double intermediate system.
We have $E^\ast$=17 MeV which corresponds to the average level spacing
$D\sim 10^{-11}$ MeV and the total width for evaporation from
the excited double intermediate system $\Gamma\leq 0.1$ keV (see Fig. 7
in Ref. [4]). Accordingly, the average time it takes for the hot
intermediate system to evaporate one nucleon is about $6\times 10^{-18}$
sec. This corresponds to $\sim 2000$ complete revolutions of the 
intermediate system with $\hbar\omega=1.2$ MeV. This is about 
three orders of magnitude larger than a typical average number ($\sim 1-3$)
of complete revolutions of the hot double intermediate system 
before its disintegration into two fragments [7,8,9]. This indicates that
the production of the projectile like ejectiles 
is a primary binary process
which is not affected by the nucleon evaporation from the hot deformed
intermediate system.

{\bf IV. Tests of a statistical significance of the data}

One possibility to find out if the oscillations in the individual 
excitation functions 
measured for each of the two different targets (Fig. 6) are true
oscillations is to calculate the experimental normalized variances of the 
oscillations, $C(\varepsilon =0)$. Here $C(\varepsilon)
=<\Delta\sigma (E+\varepsilon )\Delta\sigma (E)>$ is a cross section 
energy autocorrelation function, $\Delta\sigma (E)=
(\sigma(E)/<\sigma(E)>-1)$ is a relative oscillating yield, and 
$<\sigma(E)>$ is an energy averaged smooth cross section which was
obtained from the best second order polynomial fit of the original data.
For the two independent measurements of the N oscillating yields
(Fig. 6) at $\theta=53^\circ$ we obtain $C(\varepsilon=0)=0.015\pm 0.0035$
for the first target and  $C(\varepsilon=0)=0.017\pm 0.004$ for 
the second one, where the uncertainties are due to the finite data range
only [5]. For the O oscillating yields
 at $\theta=53^\circ$ we have $C(\varepsilon=0)=0.012\pm 0.003$
for the first target and  $C(\varepsilon=0)=0.016\pm 0.0037$ for 
the second one.
This is to be compared with the quantities $1/N$, which represent 
$C(\varepsilon =0)$ corresponding only to statistical uncertainties due 
to the finite average counting rate $N$. For the N yield we have 
$1/N=0.004$ and for the O yield $1/N=0.0035$.  
Therefore, for $\theta=53^\circ$,
the experimental values of $C(\varepsilon=0)$ are  
larger by a factor of $\sim 3$ than $1/N$ expected based on finite statistics. 
Similarily, for the two independent measurements of the N oscillating yields
(Fig. 6) at $\theta=38^\circ$ we obtain $C(\varepsilon=0)=0.0024\pm 0.0006$
for the first target and  $C(\varepsilon=0)=0.0028\pm 0.0007$ for 
the second one.
For the O oscillating yields
 at $\theta=38^\circ$ we have $C(\varepsilon=0)=0.0024\pm 0.0006$
for the first target and  $C(\varepsilon=0)=0.0022\pm 0.00055$ for 
the second one. These values are larger by a factor of $\sim 3$ than
corresponding average inverse counting rates ($1/N=0.0008$ for 
the N products
and $1/N=0.0007$ for the O products) at $\theta=38^\circ$.
The above analysis indicates that the 
oscillations shown in Fig. 6 are true oscillations and do not result from 
insufficient statistics.

Another indication for the statistical significance of the oscillations 
in Fig. 6 can be revealed from the analysis of probability
distributions of the properly scaled cross section relative deviations, 
($\sigma_i/<\sigma_i>-1)/(1/N_i)^{1/2}$, 
from the energy smooth background $<\sigma(E)>$. Here, $\sigma_i=
\sigma(E_i)$, $<\sigma_i>=<\sigma(E_i)>$ 
is an energy averaged smooth cross section 
obtained from the best second order polynomial fit of the data,
and $N_i$ is the counting rate for the $E_i$ energy step.
Suppose that the cross section energy oscillations in Fig. 6 
are not true oscillations but originate from the finite count rate.
If this would be the case then the probability distribution of
$(\sigma_i/<\sigma_i>-1)/(1/N_i)^{1/2}$ should be a Gaussian distribution
with zero expectation and unit standard deviation (variance).
Gaussian distributions and the actual probability distributions of 
absolute values of
the measured cross section deviations from the energy smooth 
background are presented in Figs. 7, 8 and 9.
One observes that the experimental probability distributions are
systematically wider than Gaussian distribution with unit standard
deviation. Also 21$\%$ of all the deviations exceed three
standard deviations (Fig. 9).

As a first step in evaluation of the statistical significance
of the non-reproducibility (Fig. 6) we calculate correlation
coefficients, $\varrho$, between the corresponding oscillating yields 
produced with the two different target foils:
$$
\varrho=(1/n)\sum_{i=1}^n(\sigma_i^{(1)}/<\sigma_i^{(1)}>-1)
(\sigma_i^{(2)}/<\sigma_i^{(2)}>-1)/[C^{(1)}(\varepsilon=0)
C^{(2)}(\varepsilon=0)]^{1/2},
\eqno{(3)}
$$
where
$$
C^{(1,2)}(\varepsilon=0)=
(1/n)\sum_{i=1}^n(\sigma_i^{(1,2)}/<\sigma_i^{(1,2)}>-1)^2,
$$
$\sigma_i^{(1,2)}=\sigma^{(1,2)}(E_i)$, $<\sigma_i^{(1,2)}>=
<\sigma^{(1,2)}(E_i)>$, and $n$ is a number of energy steps.
Indices $(1,2)$ correspond to the first and second measurement
(target), accordingly.
We find $\varrho=0.24\pm 0.06$ for the N products at $\theta_{lab}=38^\circ$,
$\varrho=0.23\pm 0.06$ for the O products at $\theta_{lab}=38^\circ$,
$\varrho=0.09\pm 0.022$ for the N products at $\theta_{lab}=53^\circ$ and
$\varrho=0.06\pm 0.015$ for the O products at $\theta_{lab}=53^\circ$,
where the uncertainties are due to the finite data range only [5].
This indicates that the non-self-averaging non-reproducible components
of the cross sections for the two 
measurements oscillate around each other in nearly statistically 
independent uncorrelated way.

Consider a probability distribution of
$$
[\sigma_{1}(E)-\sigma_{2}(E)]/(\delta\sigma_1^2+\delta\sigma_2^2+
2\rho\delta\sigma_1\delta\sigma_2)^{1/2},
\eqno{(4)}
$$
where
$$
\delta\sigma_{1,2}^2=(1/n)\sum_{i=1}^n(\sigma_i^{(1,2)}-
<\sigma_i^{(1,2)}>)^2,
\eqno{(5)}
$$
$$
\rho=(1/n)\sum_{i=1}^n(\sigma_i^{(1)}-<\sigma_i^{(1)}>)
(\sigma_i^{(2)}-<\sigma_i^{(2)}>)/\delta\sigma_1\delta\sigma_2,
\eqno{(6)}
$$
and indices $(1,2)$ stand for the first and second measurement
(target), respectively.

Suppose that the non-reproducibility of the cross section energy 
oscillations in Fig. 6 
is not a true effect but originate from the finite count rates.
If this would be the case then the probability distribution of
the quantity (4) with 
$$
\delta\sigma_{1,2}^2=
(1/n)\sum_{i=1}^n
<\sigma_i^{(1,2)}>^2/N_i^{(1,2)},
\eqno{(7)}
$$
should be a Gaussian distribution
with zero expectation and unit standard deviation (variance). In Eq. (7)
$N_i^{(1,2)}$ are the counting rates for the $E_i$ energy step in the
first and second measurements, accordingly.
Gaussian distributions with unit standard deviation and the actual 
experimental probability distributions of 
absolute values of the quantities (4) with $\delta\sigma_{1,2}^2$
given by Eq. (7) are presented in Figs. 10 and 11.
One observes that the experimental probability distributions are
systematically wider than Gaussian distribution with unit standard
deviation. A level of the non-reproducibility exceeds three
standard deviations for 18$\%$ of all the cross section differences  
measured (Fig. 11). This indicates that the non-reproducibility
of the cross section energy oscillations (Fig. 6) measured with different 
target foils of nominally the same thickness 
is of a statistical significance.

On the contrary, the two runs for the same target foil (Fig. 5)
produce the reproducible cross section energy oscillations (see Figs. 
10 and 11).

Our data indicate a strong correlation between the N and O dissipative
yields for each of the {\sl individual} measurement (target). For example,
for  $\theta_{lab}=38^\circ$, a correlation coefficient between
the N and O cross sections is  $\varrho=0.6\pm 0.15$ for the first
target and  $\varrho=0.61\pm 0.15$ for the second one,
where the uncertainties are due to the finite data range only [5].

Such a strong
($\simeq 0.5-0.9$) correlation between strongly dissipative yields with 
different charges or  different masses is a characteristic feature
of most of the systems measured [7-9]. This suggests that our
data, for each individual measurement (target), are of a similar
character as those reported in Refs. [7-9]. A strong correlation
 between dissipative yields with 
different charges and different masses is consistent with the 
interpretation of the MC [9,14].

Measurements of the excitation function oscillations, for the single
target foil, with distinction
of isotopes for the yields from the $^{19}$F+$^{51}$V strongly dissipative
collisions were reported
in ref. [7], Wang Qi et al. (1996). These data demonstrate a strong,
$\simeq 0.8-0.9$, correlation between different isotopes. Assuming
that the present data, for each of the two individual
 measurements (targets),
are of a similar character as those reported in ref. [7], Wang Qi et al. 
(1996) one would expect that a level of non-reproducibility for different
target foils for isotopes
of N and O would be similar to that observed without distinction of
isotops (Fig. 6).

It is clear that the random matrix theory and statistical 
theory of Ericson fluctuations are of no help for the
interpretation of the experimental results reported in this paper.
Indeed, the theory of Ericson fluctuations is conceptually based on 
the statistical
model which disregards outright micro-channel correlations [3,4]. Therefore,
the necessary conditions for applicability of the random matrix theory, 
statistical model and theory of Eriscon fluctuations to the interpretation
of
the data reported here must be (i) absence of oscillations in the cross
sections, i.e. energy smooth excitation functions for each of the
individual measurement, and (ii) reproducibility of these energy smooth
cross sections in the measurements with different target foils.
Both of these conditions are not met for the
data sets reported in this paper. 

A quantitative interpretation of the energy oscillations in the individual
data sets in terms of the spontaneous self-organisation, 
non-equilibrium micro-channel correlation phase transitions and 
anomalously slow phase 
randomization in highly excited strongly interacting finite quantum many-body 
systems [15] will be presented in a separate communication.

{\bf V. Conclusion}

In conclusion, the two independent measurements with different target foils 
of nominally the same thickness
indicate statistically significant non-reproducibility of the
cross sections for the  $^{19}$F+$^{93}$Nb DHIC.
The non-reproducibility
is consistent with the recent theoretical arguments on spontaneous
coherence, slow phase randomization and anomalous sensitivity in
finite highly excited quantum systems.
If this non-reproducibility is confirmed in future experiments it will 
signal that a realization of 
Wigner's dream [10], a theory for the transition amplitude 
correlations, will require conceptual revision
of modern understanding of microscopic and mesoscopic quantum 
many-body systems. 

This work was supported by the Natural Science Foundation of 
China (N. 19775057).
The authors wish to thank the staff of the HI-13 accelerator at CIAE. 
The work by S. Kun was supported by the Max Planck Society Fellowship.
S. Kun is grateful to members of the Max-Planck-Institute f\"ur
Kernphysik, Heidelberg for a warm hospitality during his visits
on May-July 2001 and October-November 2002. We thank Christian
Beck and Lewis T. Chadderton for useful discussions and suggestions.

\vfill\eject

{\bf References}

\parindent=0pt

[1] E. Wigner, in {\sl Statistical Properties of Nuclei}, edited by 
J.B. Garg, (Plenum Press, New York-London, 1972), p. 11.

[2] P.W. Anderson, {\sl Basic Notions of Condensed Matter Physics},
Frontiers in Physics, vol. 55, (The Benjamin-Cummings,
1984), pp. 71-72.

[3] T. Guhr, A. M\"uller-Groeling, and H.A. Weidenm{\" u}ller, 
 Phys. Rep. {\bf 299}, 189 (1998), and references therein.

[4] T. Ericson and T. Mayer-Kuckuk, Ann. Rev. Nucl. Sci. {\bf 16}, 183 
(1966).

[5] W.R. Gibbs, Phys. Rev. {\bf 139}, B1185 (1965); P.J. Dallimore and
I. Hall, Nucl. Phys. {\bf 88}, 193 (1966).

[6] H.A. Weidenm\"uller, Progr. Part. Nucl. Phys. {\bf 3}, 49 (1980),
 and references therein.

[7] G. Pappalardo, Nucl. Phys. A{\bf 488}, 395c (1988);
A. De Rosa {\sl et al.},
 Phys. Rev. C{\bf 37}, 1042 (1988);
{\sl ibid.}   C{\bf 40}, 627 (1989);
{\sl ibid.} C{\bf 44}, 747 (1991);
 Wang Qi {\sl et al.},
Chin. J.  Nucl. Phys.  {\bf 15}, 113 (1993);
F. Rizzo {\sl et al.},
 Z. Phys. A{\bf 349}, 169 (1994);
M. Papa {\sl et al.}, Z. Phys. A{\bf 353}, 205 (1995);
Wang Qi {\sl et al.},
Phys. Lett. B{\bf 388}, 462 (1996);
I. Berceanu {\sl et al.}, Phys. Rev. C{\bf 57}, 2359 (1998);
M. Papa {\sl et al.},
Phys. Rev. C{\bf 61}, 044614 (2000).

[8] T. Suomij\"arvi {\sl et al.}, 
Phys. Rev. C{\bf 36}, 181 (1987).

[9] S.Yu. Kun {\sl et al.},
Z. Phys. A{\bf 359}, 263 (1997).

[10] E. Wigner, in reference [1], pp. 621,622,635,636.

[11] S.Yu. Kun,  Z. Phys. A{\bf 357}, 255 (1997).

[12] S.Yu. Kun, Z. Phys. A{\bf 357}, 271 (1997).

[13] S.Yu. Kun and A.V. Vagov, Z. Phys. A{\bf 359}, 137 (1997).

[14] S.Yu. Kun, Phys. Rev. Lett. {\bf 84}, 423 (2000).

[15] S.Yu. Kun, in {\sl Non-Equilibrium and Nonlinear Dynamics in Nuclear 
and Other Finite Systems}, edited by Zhuxia Li, Ke Wu, Xizhen Wu, Enguang 
Zhao and Fumihiko Sakata, the American Institute of Physics Proc.
Series {\bf 597}, 319 (2001).

\vfill\eject

{\bf Figure Captions}

Fig. 1.
$\Delta E-E$ scatter-plots obtained in the  $^{19}$F+$^{93}$Nb
dissipative heavy-ion collision at 
$\theta_{lab}=38^\circ$ (left panel) and 
$\theta_{lab}=53^\circ$ (right panel) at 
$E_{lab}=100.25$ MeV.
The Fig. also shows energy gates used in the analysis.

Fig. 2.
$\Delta E-E$ scatter-plots obtained in the  $^{19}$F+$^{12}$C
reaction at $\theta_{lab}=38^\circ$ (left panel) and 
$\theta_{lab}=53^\circ$ (right panel) at $E_{lab}=100.25$ MeV.

Fig. 3.
Energy spectrum of Z=7 dissipative fragments produced in the  
$^{19}$F+$^{93}$Nb dissipative heavy-ion collision 
at $\theta_{lab}=38^\circ$ and $E_{lab}=103.25$ MeV for one run
in the second experiment (see Fig. 5).

Fig. 4.
Angular distributions of N dissipative yield of the  
$^{19}$F+$^{93}$Nb dissipative heavy-ion collision 
at $E_{lab}=100$ MeV and 105 MeV. The solid and dashed lines are
for the eye guide.

Fig. 5.
Excitation functions for the N and O yields of the $^{19}$F+$^{93}$Nb
strongly dissipative heavy-ion collisions obtained in the two 
runs (triangles and crossed circles) for the same 
single target foil in the second experiment. 
The error bars are statistical only.

Fig. 6.
Excitation functions for the N and O yields of the $^{19}$F+$^{93}$Nb
strongly dissipative heavy-ion collisions obtained in the two independent
experiments. Full dots correspond to the first experiment 
and open squares to the second one. The error bars are statistical only.

Fig. 7.
Probability distributions of absolute values of the cross section 
relative deviations from the energy smooth background obtained in the first
measurement (dashed hystograms). Solid hystograms are
Gaussian distributions with unit standard deviation expected based
on the finite count rates only (see text).

Fig. 8.
The same as in Fig. 7 but for the second measurement with different 
target (see text).

Fig. 9.
The same as in Figs. 7 and 8 but for the sum of all 8 sets of the individual
probability distributions from Figs. 7 and 8.

Fig. 10.
Probability distributions of absolute values of the properly scaled
differences between the cross sections obtained in the two measurements
with different target foils (dashed hystograms). Dotted hystograms
are probability distributions of absolute values of the properly scaled
differences between the cross sections obtained in the two runs 
with the same target foil for the second measurement.
Gaussian distributions (solid hystograms) with unit standard deviation 
expected based on the finite count rates only (see text).
 
Fig. 11.
The same as in Fig. 10 but for the sum of all 4 sets of the individual
probability distributions from Fig. 10.

\vfill\eject\end